\documentstyle[preprint,aps,prl,epsf]{revtex}

\begin{document}

\preprint{$\begin{array}{l}
\hbox{SHEP\  93/94-19} \\
\hbox{hep-ph/9406303}
\end{array}$}

\vspace{0.2in}

\title{Unification Constraints in the Next-to-Minimal
Supersymmetric Standard Model}

\vspace{0.2in}

\author{T. Elliott, S.F. King
\footnote{Email \quad {\tt king@soton.ac.uk}}
\\
 {\it Physics Department, University of Southampton,
\\ Southampton SO9 5NH, UK.\\}
\vspace{0.2in}
P.L. White\footnote{Email \quad {\tt plw@thphys.ox.ac.uk}}
\\
{\it Theoretical Physics, University of Oxford, \\
1 Keble Road, Oxford OX1 3NP, UK.\\ }
}

\vspace{0.2in}

\date{\today}

\maketitle

\begin{abstract}

We perform a renormalisation group analysis of the next-to-minimal
supersymmetric standard model
(NMSSM) based on the following constraints:
two-loop gauge coupling unification at a variable scale $M_X$, running
the gauge couplings through the low energy thresholds; universal soft
supersymmetry breaking parameters; correct electroweak symmetry
breaking. The phenomenological implications
of our results include a standard model-like
Higgs boson of mass in the range 70-140 GeV.

\end{abstract}


%\narrowtext

Supersymmetry (SUSY) \cite{SUSY} is an attractive candidate for new
physics beyond the standard model at the TeV scale. The existence of
SUSY would greatly ease the naturalness problems associated with the
construction of a  grand unified theory (GUT) of the strong and
electroweak interactions. The basic idea of GUTs is that the gauge
couplings, which govern the strength of the strong and electroweak
interactions at low energy, are actually equal to some unified coupling
$g_X$ at some very high scale $M_X$ due to their renormalisation group
(RG) running \cite{GQW}.  The original motivation for SUSY broken at a
TeV was to help to stabilise the Higgs mass against GUT scale quadratic
radiative corrections. The combination of SUSY and GUTs has recently
found some indirect experimental support due to the accurate
measurement of the strong and electroweak couplings  on the Z pole by
LEP.  These measurements are inconsistent with gauge coupling
unification if a standard model desert is assumed, but are consistent
with unification if a SUSY desert above the TeV scale is assumed
\cite{Amaldi}.

In the first post-LEP analyses \cite{Amaldi} the whole SUSY spectrum
was either assumed to be degenerate at the scale $M_{SUSY}$, or smeared
around this scale. In reality the SUSY partners may have a complicated
spectrum, parametrised by a large number of soft SUSY-breaking
parameters, which may spread over one or two orders of magnitude of
masses. In order to reduce the number of independent soft SUSY-breaking
parameters one may appeal to supergravity or superstring scenarios
which involve the notion of SUSY breaking in a hidden sector coupled
only gravitationally to our observable sector. By this means, in the
minimal supersymmetric standard model (MSSM), one ends up with just
four independent soft SUSY breaking parameters: $m_0$, $M_{1/2}$, $A_0$
and $B_0$ corresponding to the universal soft scalar, gaugino,
trilinear and bilinear couplings, respectively. Using these four
parameters, together with the top quark Yukawa coupling which plays an
important role in driving electroweak symmetry breaking \cite{RR},
several groups have performed an RG analysis whose goal is to predict
the SUSY and Higgs spectrum  and then use this spectrum as the basis of
a more reliable estimate of gauge coupling unification by running the
gauge couplings through the various thresholds from the Z mass $M_Z$ up
to a unification scale $M_X$ \cite{RR,all}. Apart from the constraints
of gauge coupling unification and correct electroweak symmetry
breaking, there are  various other phenomenological and cosmological
constraints  which may be applied and recent studies have  concluded
that it is possible to satisfy all these constraints simultaneously
\cite{constraints}.

Here we shall consider a slightly different low
energy SUSY model, but one which is equally consistent with
gauge coupling unification, namely the so-called next-to-minimal
supersymmetric standard model (NMSSM) \cite{NMSSM1,NMSSM2,NMSSM3}.
The basic idea of the NMSSM is to add just one extra gauge singlet
superfield $N$ to the spectrum of the MSSM, and to
replace the $\mu$-term in the
MSSM superpotential with a purely cubic superpotential,
\begin{equation}
\mu H_1H_2 \longrightarrow \lambda NH_1H_2 - {k\over 3} N^3.
\end{equation}
The motivation for this ``minimal non-minimal'' model is
that it solves the so-called
$\mu$-problem of the MSSM \cite{mu}
in the most direct way possible by
eliminating the $\mu$-term altogether,
replacing its effect by the vacuum expectation value (VEV)
$<N>=x$, which may be naturally related to the usual Higgs VEVs
$<H_i>=\nu_i$.
However there are
other solutions to the $\mu$-problem \cite{mu}. Also, the  inclusion of
singlets may cause the destabilisation of the hierarchy if there are
strong couplings to super-heavy particles such as Higgs colour triplets
\cite{stabone}.  Recently similar effects have been shown to result
from non-renormalisable operators suppressed by powers of the Planck
mass  \cite{stabtwo}. The dangerous non-renormalisable operators
would require that gravity violate the $Z_3$ symmetry which is
respected by
the renormalisable operators of the theory. Our view is that since
these effects are model-dependent the NMSSM
is as well motivated as the MSSM and should be studied to the same
level of approximation. Only by so doing may the two models be
phenomenologically compared when (or if) Higgs bosons and SUSY particles
are discovered.

In this Letter we present our first results of  a GUT-scale RG analysis
of the NMSSM, imposing simultaneously the constraints of correct
electroweak symmetry breaking and coupling constant unification. In our
analysis we shall drop all quark and lepton Yukawa couplings apart from
that of the top quark $h_t$. We shall assume a soft SUSY breaking
potential which  gives rise to three universal soft parameters at the
$M_X$, namely $m_0$, $M_{1/2}$, $A_0$, mentioned above \footnote{ Note
that the $B$ parameter does not occur in the NMSSM since we have set
$\mu=0$.}. We shall use the two-loop gauge and Yukawa SUSY RG equations
and one-loop soft SUSY RG equations to integrate
from  the GUT scale $M_X$ to a low energy
renormalisation scale $Q$. Below $M_X$, the
three soft mass parameters evolve into 32 separate soft parameters,
corresponding to 3 gaugino masses $M_i$, 11 trilinear couplings $A_i$
(those of the MSSM plus $A_{\lambda}$ and $A_k$) and 18 scalar masses
$m_i^2$ (including the soft Higgs masses $m_{H_i}^2$, $m_N^2$). We
assume no inter-generational mixing, and since we drop Yukawa couplings
which are small compared to $h_t$, the first and second generation soft
masses will run identically. The unification constraints are:
$g_i^2(M_X) = g_X^2$,  $ \lambda (M_X)  =  \lambda_0$, $  k(M_X)  =
k_0$,  $ h_t(M_X)  = {h_t}_0$, $M_i(M_X) =  M_{1/2}$,  $m_i^2(M_X) =
m_0^2$,  $A_i(M_X)  =  A_0$.

In practice we shall scale all dimensionful quantities by appropriate
powers of $M_{1/2}$, and denote the resulting dimensionless quantities
by a tilde.
At the low energy $\overline{MS}$ scale $Q$
we find the global minimum of
the one-loop effective neutral Higgs potential,  scaled by $M_{1/2}$,
including loops of top quarks and stop squarks\cite{rad4,ekw}.
We only
consider the neutral scalar Higgs potential, since there are simple
analytical conditions which test for the presence of squark and slepton
VEVs \cite{NMSSM2}.  Although these are not guaranteed to prevent all
possible charge or colour breaking minima as discussed by Gunion {\it
et al} in reference \cite{NMSSM2}, they are sufficient to avoid the
most dangerous case of slepton VEVs, while a complete analysis of all
possible colour breaking minima in the one loop effective potential is
not feasible. We implement these conditions after running to an energy
scale of order a typical slepton VEV. In addition we ensure that the
mass squared of the charged Higgs remains positive in order to avoid
breaking electromagnetism. The result of minimisation yields the scaled
VEVs $\tilde{\nu}_i, \tilde{x}$, from which the condition $\nu=174$ GeV
fixes $M_{1/2}$.  Given $M_{1/2}$ this enables all the soft parameters
and the VEVs to be determined and,
as described elsewhere \cite{rad4,ekw},
it is then straightforward to calculate the top and stop corrections to
the Higgs boson mass spectrum.

The SUSY spectrum is calculated at
tree-level using standard results \cite{SUSY}, apart from squark masses
which  we calculate using the one-loop effective potential.  There are
a number of phenomenological  constraints which will cut down the
allowed regions of parameter space.  We shall require that all sleptons
and stops are heavier that 43 GeV and all charginos are heavier than 47
GeV.  Charged Higgs bosons are required to be heavier than 45 GeV.
Gluinos and squarks other than stops  are required to be heavier than
100 GeV. The lightest neutralino is required to be the lightest SUSY
particle. The lightest CP-even neutral Higgs boson $h$ is required to
satisfy $m_h/R^2_{ZZh} \geq 60$ GeV, where $m_h$ is the mass and
$R_{ZZh}$ is the $ZZh$ coupling scaled by the standard model coupling.
We shall require a large top quark mass, $m_t>150$ GeV \cite{CDF},
which puts a strong
restriction on the parameter space of this model. In our analysis we
input the gauge couplings $g_1(M_Z)$ and $g_2(M_Z)$ and run them up
through the 24 SUSY and Higgs thresholds to find $M_X$ and $g_X$, which
must of course be consistent with our original input values. By
iterating this procedure we obtain solutions which satisfy both the
requirements of correct electroweak symmetry breaking and coupling
constant unification {\it simultaneously}. According to our procedure
the parameters $M_X$ and $g_X$ are determined from $g_1(M_Z)$ and
$g_2(M_Z)$, and so our input parameter set is: $\lambda_0$, $k_0$,
${h_t}_0$, $\tilde{m}_0^2$, $\tilde{A}_0$, with the value of $M_{1/2}$
determined from the VEV. The physically relevant quantities
$\alpha_s(M_Z)={g_3}^2(M_Z)/4\pi$, $m_t$, $r=x/{\nu}$ and $\tan
\beta=\nu_2/\nu_1$ are all outputs \footnote{Our quoted values of $m_t$
always refer to the one-loop  physical pole mass $m_t=m_t^{pole}=
m_t(m_t)[1+\frac{4}{3\pi}\alpha_s(m_t)]$. Also, a word about our
conventions is in order. In the convention of Eq.(1), we input positive
values of $\lambda_0$, $k_0$, and both positive and negative values of
$A_0$, and look for solutions with both positive and negative  VEVs,
except that we require $\nu_2>0$. Such solutions can always be
re-interpreted in terms of purely positive VEVs using the three
symmetries of the one-loop effective potential: (i) $\lambda
\rightarrow - \lambda , k \rightarrow -k, x\rightarrow -x$, (ii)
$\lambda \rightarrow -\lambda, \nu_1 \rightarrow - \nu_1$, (iii) $\nu_1
\rightarrow - \nu_1, \nu_2 \rightarrow -\nu_2$.}.

A somewhat similar analysis to that described has been performed in the
NMSSM by another group \cite{ell,ell2,ell3}.  We have properly included
the low energy threshold effects in our unification analysis, whereas
in ref.\cite{ell,ell2,ell3} such effects were ignored. Since our choice
of $\alpha_3(M_Z)$ and of $M_X$ and $g_X$ are determined  by the
requirements of consistent unification and thus vary with differing
spectra, our approach is consequently much more computationally
intensive since we have to run down the couplings, minimise the
potential, calculate the spectrum, then run the couplings up through
the thresholds, iterating this procedure many times in order to obtain
a single point in parameter space which is consistent with unification
and all the other constraints.

We initially considered a grid of 360,000  values of parameters in the
ranges: $\lambda_0=0.01-2.0$, $k_0=0.01-2.0$, ${h_t}_0=0.5,1,2,3$,
$\tilde{m}_0=0.2-5.0$,  ${A}_0/m_0=$-4 to +4, then later investigated
the phenomenologically interesting regions in more detail. For the
acceptable regions of parameter space we find
that $m_t>150$ GeV is attainable for
$k_0<\lambda_0<h_{t0}$, with $\lambda_0<1$ and
$0.5\stackrel{<}{\sim} h_{t0}\stackrel{<}{\sim}3$,
where the upper limit on $h_{t0}$ is a requirement of perturbation
theory.
Although all values of $h_{t0}$ in this range lead to consistent
solutions with $|{A}_0/m_0|\approx 3-4$, for
$h_{t0}\stackrel{<}{\sim}0.5$ we also find solutions for smaller
$|A_0/m_0|$, down to zero.
However the cut on the top mass of $m_t>150$ GeV excludes
this region, and so we discuss it no further here.
The successful values of $\tilde{m}_0$ vary across the whole range, but
are correlated with $h_{t0}$ and ${A}_0/m_0$. To be specific, we find
for $|{A}_0/m_0|\approx 3$ that $h_{t0}=0.5,1,3$,  is associated with
$\tilde{m}_0\sim 0.5-1,1-2,2-5$,
while smaller values of $\tilde m_0$ can
be obtained for larger $|A_0/m_0|$, although
the maximum value of  $|A_0/m_0|$ is about 4 (because of
the slepton VEVs constraint). There are
also correlations
between $\lambda_0$ and
$k_0$ at the unification scale, as discussed later.
In the successful regions of parameter space we find
$|\tan \beta| \sim 3-20$ (or
larger, but our approximations break down here), while $|r|$ is
virtually directly proportional to $M_{1/2}$, with $|r|\gg 1$ even for
small $M_{1/2}$.
Thus, as in ref.\cite{ell}, we find that
the allowed region of parameter space is characterised by small
dimensionless couplings $|\lambda_0|<1$, $|k_0|<1$ and large singlet
VEV $|x|>1000$ GeV, corresponding to the approximate MSSM limit of the
model \cite{NMSSM3}.

To illustrate the correlation between $\lambda_0$ and
$k_0$, in Fig.1 we show contours of $h_{t0}$ in the $\lambda_0-k_0$
plane with $\tilde{m}_0=2$, $M_{1/2}=500$ GeV and $|A_0/m_0|=3$.
Although we could have chosen other values of
$M_{1/2}$, the qualitative features of Fig.1 would be
unaltered, and the value of $k_0$ would only change by a few per cent,
since in this region the value of $M_{1/2}$ is a very sensitive
function of $\lambda_0$ and $k_0$.
The $h_{t0}=1,2,3$ contours correspond to $m_t\sim 175, 185, 190$ GeV,
respectively, with $|\tan\beta|$ varying from 3-7 and $|r|$ from
100-30, from the lowest to the highest value of $k_0$ respectively.
Values of $k_0$ beyond the ends of the plot are forbidden by the
requirements of electroweak symmetry breaking, although the exact
ranges of acceptable $k_0$ could be altered by varying our choice of
$A_0$ and $\tilde m_0$ slightly. Similarly, $m_t$ would be altered if
we use a different $M_{1/2}$ primarily due to the resulting change in
$\alpha_s(M_Z)$. As observed by other authors \cite{Amaldi,RR} we find
that larger values of $M_{1/2}$ result in smaller values of
$\alpha_s(M_Z)$.
It should be noted that although changing the sign of $A_0$ leads to
roughly the same output data with a change in the sign of $r$, this is
only approximate, and in particular one sign of $A_0$
leads to a less restrictive slepton VEV
constraint.

We find that, for many choices of our input parameters, there appears
to be a certain minimum value of $M_{1/2}$ (recall that this is an
output parameter).  For example, if we impose a cut of
$M_{1/2}\stackrel{<}{\sim} 150$ GeV, then $m_t>150$ GeV is only
possible for $h_{t0}\approx 1.5-3$, $|A_0/m_0|\approx 3$,
$\tilde{m}_0\approx 3-5$, corresponding to $m_t\sim 185-195$ GeV.
The origin of the minimum value of $M_{1/2}$ is due to a different
competing global minimum of the one-loop effective potential, which
becomes preferred below a minimum $M_{1/2}$ value. This happens because
the logarithm in the one-loop effective potential depends both on the
renormalisation scale $Q$
(which we have so far taken to be 150 GeV) and on $M_{1/2}$.
To test the reliability of our results, we have investigated their
$Q$-dependence. In doing this we have consistently used the same value
of $Q$ both in the RG running of masses and couplings and in the
one-loop effective potential, since there is a significant cancellation
between the effects of changing $Q$ in these two places.

Although varying $Q$ leads to very little
change in the VEVs for any given minimum, it can alter the VEVs
substantially by changing which of the minima is deepest (there are
often minima with very different VEVs but with similar values of the
potential). Given this behaviour, one must be very careful about
drawing overly restrictive conclusions based on which of the various
minima is preferred for a given region of parameter space.
We find that reducing $Q$
will typically lead to a
reduction of the minimum value of $M_{1/2}$
in those regions where small $M_{1/2}$ is not possible
for $Q=150$ GeV, but does not
otherwise qualitatively change the behaviour.
Note that
the range of parameters  $h_{t0}\approx
1.5-3$, $|A_0/m_0|\approx 3$, $\tilde{m}_0\approx 3-5$
appears to be a
``safe'' range from the point of view of achieving small values of
$M_{1/2}\stackrel{<}{\sim}150$GeV independently of the choice of $Q$.
This dependence on $Q$ is an
inherent problem in this model because of the
form of the effective potential which has several competing minima.
In both our analysis and that of ref.\cite{ell,ell2,ell3}
electroweak symmetry breaking has been analysed using the
one-loop effective potential, including top quarks and stop squarks.
In principle the $Q$ dependence could be eliminated by including all
particles in the effective potential, and possibly working beyond
one-loop. Here we shall restrict ourselves to a discussion of the
$Q$-dependence of our results.

We now consider two examples of the sort of Higgs and SUSY spectrum
one might typically expect in the constrained NMSSM, the first with
relatively large values of $h_{t0}$ and $\tilde{m}_0$, and the second
with relatively small values of $h_{t0}$ and $\tilde{m}_0$. These two
extreme cases span the complete range of spectra predicted by the
constrained NMSSM, subject to the constraint of a large top mass
$m_t>150$ GeV.

In Fig.2 we plot the spectrum as a function of $M_{1/2}$ for the case
$h_{t0}=2$, $|A_0/m_0|=3$, $\tilde{m}_0=5$, $\lambda_0=0.4$.  With
$Q=150$ GeV we find that $M_{1/2}$ may go down to small values, which
is not unexpected since these parameters are in the ``safe'' range
$h_{t0}\approx 1.5-3$, $|A_0/m_0|\approx 3$, $\tilde{m}_0\approx 3-5$,
referred to above. Note that the experimental constraint that
the charginos are heavier than 47 GeV implies that
$M_{1/2}\stackrel{>}{\sim}70$ GeV in this case.
For $M_{1/2}\approx 100 (1000)$ GeV,  which is
controlled by choosing $k_0=0.275(0.300)$, we find $|\tan\beta |
\approx 6(8)$, $|r| \approx 13(120)$, $\alpha_s(M_Z)\approx
0.121(112)$, $M_X=2.1(0.99)\times 10^{16}$ GeV,
$g_X=0.71(0.69)$.
The lightest CP-even Higgs boson has standard model-like couplings
($R_{ZZh}>0.99$ everywhere)
and for $M_{1/2}\stackrel{<}{\sim}100$ GeV is in the LEP2 range, as are
the lighter chargino and neutralinos
which have a much stronger $M_{1/2}$ mass
dependence. The top quark mass ranges from $m_t=193-184$ GeV, being
smaller for larger $M_{1/2}$ due to $\alpha_s(M_Z)$ being consequently
smaller. For $M_{1/2}\stackrel{<}{\sim}100$ GeV,
the lightest stop and gluino are not too
much heavier than the top quark, although the remaining
sparticles and Higgs bosons are significantly
heavier than the top quark.

In Fig.3 we plot the spectrum as a function of $M_{1/2}$ for the case
$h_{t0}=0.5$, $|A_0/m_0|=4$, $\tilde{m}_0=0.5$, $\lambda_0=0.1$.
These parameters are
outside the ``safe'' range $h_{t0}\approx 1.5-3$,
$|A_0/m_0|\approx 3$, $\tilde{m}_0\approx 3-5$, in which small
$M_{1/2}$ can be achieved independently
of the choice of the renormalisation scale
$Q$, and so we plot the spectrum
for two choices of $Q$.
The main effect of changing $Q$ is to
change the value of $M_{1/2}$ at which the data cuts out.
For $Q=150(25)$ GeV, we find the minimum values
$M_{1/2}=300(125)$ GeV corresponding to $m_0=150(62.5)$ GeV.
As in Fig.2, the lightest CP-even Higgs boson
has standard model-like couplings, and for a given $M_{1/2}$ is even
lighter than in Fig.2.
For the smaller $M_{1/2}$ values which we can obtain with
$Q=25$ GeV the lighter chargino
and neutralinos may be in the LEP2 range.
Unlike Fig.2, the left-handed sleptons are now much
lighter (due to the smaller value $\tilde{m}_0=0.5$) while the lighter
stop is much heavier (due to the smaller value $h_{t0}=0.5$). The top
quark mass has a maximum value of
$m_t=175$ GeV for $M_{1/2}=125$ GeV and here $|\tan \beta |=3.4$.
The gluino in Fig.3 is now the heaviest
sparticle, whereas in Fig.2 it was
one of the lighter ones. In general
the Higgs and sparticle masses in Fig.3
are focussed into a narrower band of
masses than in Fig.2, which is a simple
result of having $\tilde{m}_0=0.5$ rather than $\tilde{m}_0=5$.

Finally it is worth comparing our
spectra to the spectra discussed in a
paper which appeared just as this present
article was being finalised
\cite{ell3}. In this most recent analysis
the Higgs and SUSY spectrum
was presented as as series of scatter
plots for masses and couplings. Most
of this data and all of ours corresponds to one of the  CP-even Higgs
bosons being almost pure gauge singlet.
Such a decoupled Higgs boson might be
expected since the constrained NMSSM is close to the MSSM limit.
In the analysis of ref.\cite{ell3}
the decoupled Higgs boson may have a mass
either less than or greater than the lighter of the other two CP-even
Higgs bosons and when the would-be
decoupled Higgs is close in mass to the lighter physical
Higgs, strong mixing can occur, leading to
two weakly coupled Higgs bosons
of the kind discussed in ref.\cite{ekw}.
However, given the range of parameters considered in this paper,
we find that the decoupled CP-even
Higgs boson is always substantially
heavier than the lighter of the two physical
CP-even Higgs bosons. Similarly the CP-odd Higgs bosons are much
heavier than the lightest CP-even Higgs boson.
The reason for this difference is simply that in ref. \cite{ell3} the
range of parameters considered exceeds the range considered in this paper.
It turns out
\footnote{U. Ellwanger, private communication} that in order to bring
down the mass of the CP-even singlet sufficiently one requires
$\tilde{m}_0\ll 0.2$, $\lambda_0< 0.01$ and $k_0\ll \lambda_0$,
corresponding to extremely large values of $r\gg 100$. We have
checked that, in this region of parameter space,
the singlet CP-even Higgs boson does indeed
become much lighter, leading to the strong mixing effect mentioned above.
However, for the range of parameters considered in the present paper,
a standard model-like Higgs boson with a mass in the range 70-140 GeV
is preferred.

\vspace{0.5in}

{\bf Acknowledgement}

We are very grateful to Ulrich Ellwanger for his help in correcting
an error in an earlier version of this paper.

\newpage

\begin{center}
{\Large \bf Figure  Captions}
\end{center}

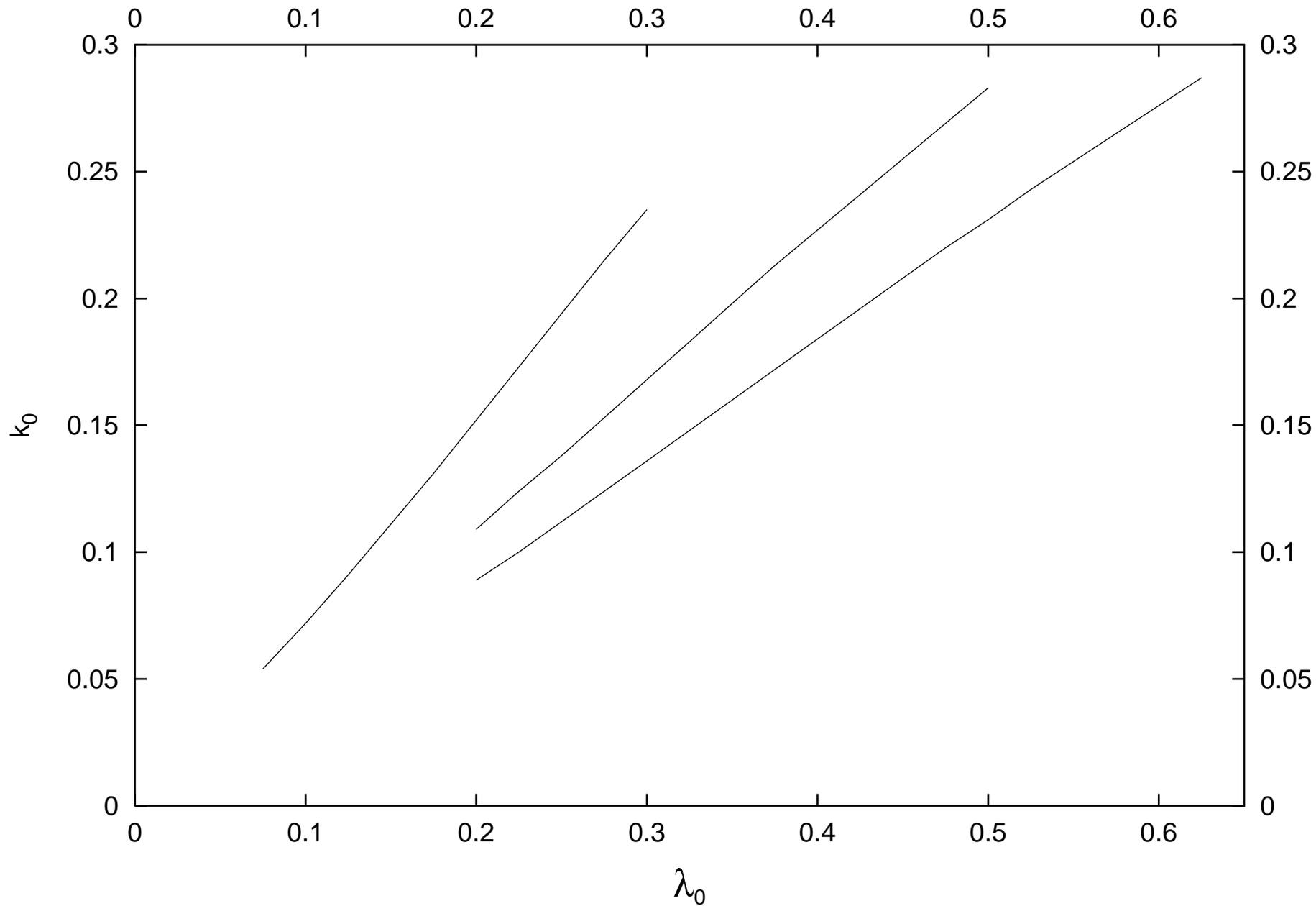
{\bf Figure 1:} A contour plot showing allowed values of
$\lambda_0$ and $k_0$ for given values of
$h_{t0}=1,2,3$  (going from left to right) in the
$\lambda_0$-$k_0$ plane, corresponding to $M_{1/2}=500$ GeV,
$\tilde{m}_0=2$ with
$|A_0/m_0|=3$.

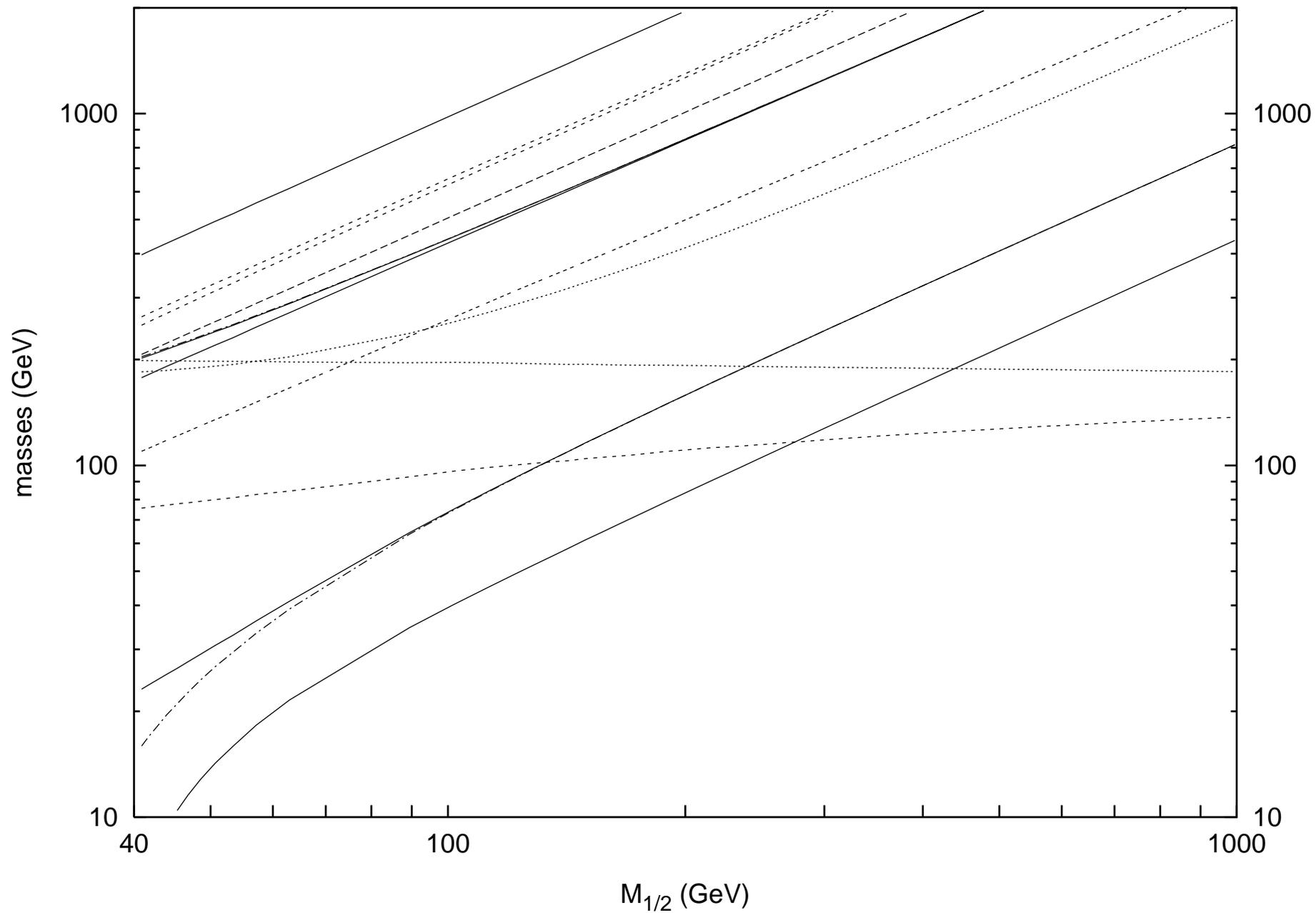
{\bf Figure 2:}
Masses of particles as a
function of $M_{1/2}$.
The input parameters are $h_{t0}=2$,
$|A_0/m_0|=3$, $\tilde{m}_0=5$, $\lambda_0=0.4$,
$k_0=0.275-0.3$. We use $Q=150$ GeV.
Neutralinos (solid lines),
charginos (dot-dashed lines),
CP-even Higgs (short dashed lines),
lighter stop and top quark
(both dotted lines), left-handed sleptons (long dashed lines),
and gluino (quadruple dashed lines)
are displayed.
One of the CP-odd Higgs and the
charged Higgs bosons (not shown)
are roughly degenerate with heavier CP-even Higgs bosons.
The remaining CP-odd Higgs boson (not shown)
is heavier than the heaviest neutralino.
The lightest CP-even Higgs
(which has standard model-like couplings),
lighter chargino and lightest two neutralinos
are all in the LEP2 range for
$M_{1/2}\stackrel{<}{\sim}100$ GeV.

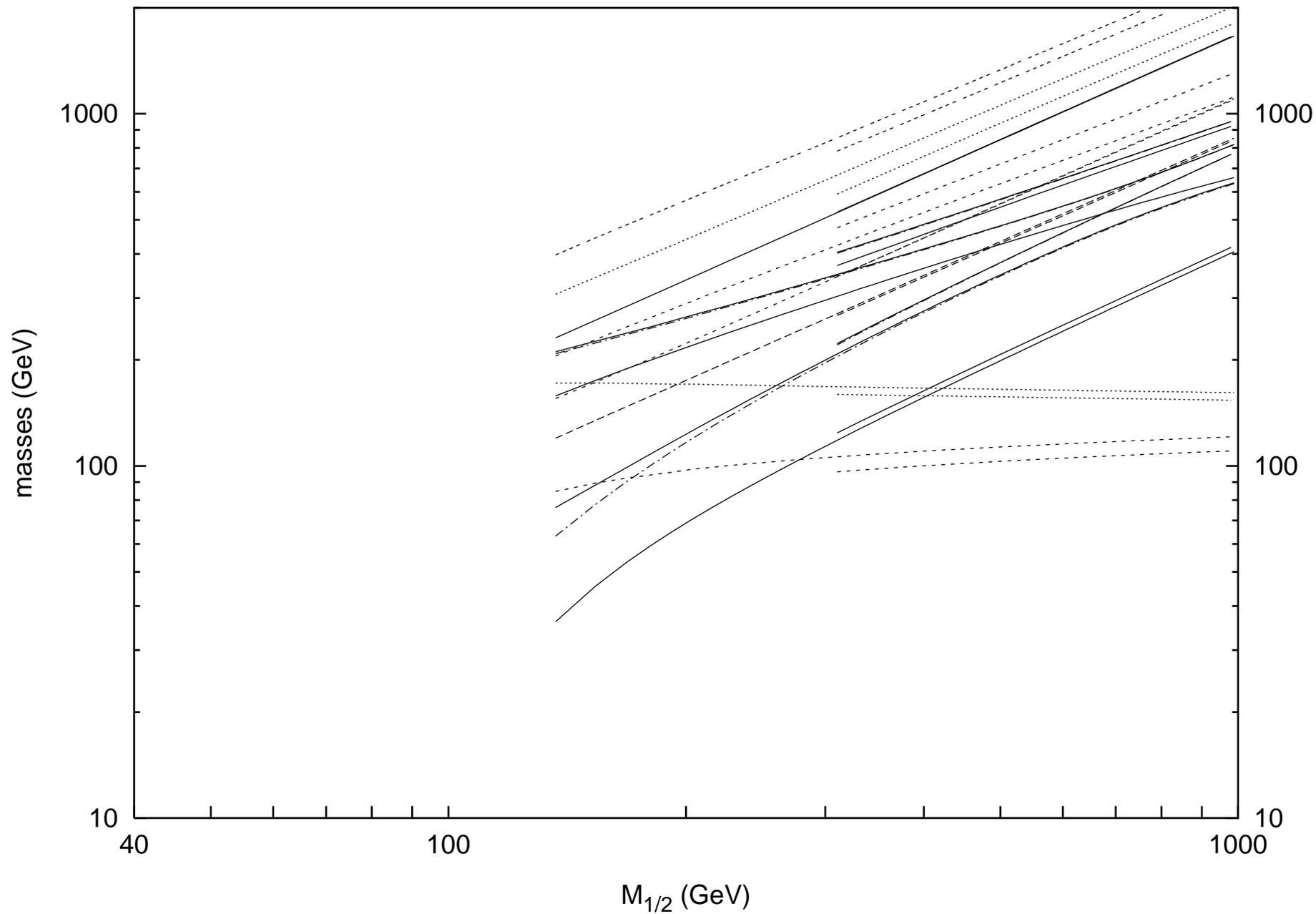
{\bf Figure 3:}
Masses of particles as a
function of $M_{1/2}$.
The input parameters are $h_{t0}=0.5$,
$|A_0/m_0|=4$, $\tilde{m}_0=0.5$,
$\lambda_0=0.1$, $k_0=0.07-0.13$.
In addition to using
$Q=150$ GeV, as in Fig.2, we also show results for
$Q=25$ GeV, corresponding to
minimum values of $M_{1/2}=300$ GeV and
$M_{1/2}=125$ GeV, respectively.
Neutralinos (solid lines),
charginos (dot-dashed lines), CP-even Higgs (short dashed lines),
lighter stop and top quark
(both dotted lines), left-handed sleptons (long dashed lines),
and gluino (quadruple dashed lines)
are displayed.
One of the CP-odd Higgs (not shown)
is roughly degenerate with the second heaviest CP-even Higgs boson.
The remaining CP-odd Higgs boson (not shown)
is roughly equal to the lighter stop mass.
The lightest CP-even Higgs (which has standard
model-like couplings as in Fig.2)
has a mass $\stackrel{<}{\sim}100$ GeV.

\end{document}